# FRONT END AND HFOFO SNAKE FOR A MUON FACILITY*

D. Neuffer[#], Y. Alexahin, Fermilab, Batavia, IL 60510, USA

## Abstract

A neutrino factory or muon collider requires the capture and cooling of a large number of muons. Scenarios for capture, bunching, phase-energy rotation and initial cooling of μ's produced from a proton source target have been developed, for neutrino factory and muon collider scenarios. They require a drift section from the target, a bunching section and a φ-δE rotation section leading into the cooling channel. The currently preferred cooling channel design is an "HFOFO Snake" configuration that cools both $\mu^+$ and $\mu^-$ transversely and longitudinally. The status of the design is presented and variations are discussed.

## INTRODUCTION

Scenarios have been developed for using muons in a storage ring based neutrino source or "neutrino factory" and in a high-energy high-luminosity "muon collider". [1, 2] The scenarios are outlined in figure 1. In both scenarios high intensity proton bunches from a proton source strike a production target producing secondary particles (mostly $\pi^{\pm}$'s). The π's decay to μ's and the μ's from the production are captured, bunched, cooled and accelerated into a storage ring for neutrinos or high energy collisions. The present paper discusses the section of the scenarios labelled the "front end" in figure 1, between the target and the accelerator for the neutrino factory and between the target and "6-D" cooling section of the muon collider.

In the Front End, pions from the production target decay into muons and are focused by magnetic fields and bunched by time-varying electric fields into an initial cooling system that forms the muons into a beam suitable for the following acceleration and/or cooling.

μ's from the target and decay are produced within a very broad energy spread and length. Initially, capture within a single bunch was considered, but that requires either very low frequency rf (<20MHz) or novel induction linacs. The scale and cost of such a system would be uncomfortably large. Instead a novel system of higher frequency rf cavities (~200—500 MHz) was developed that forms the μ's into a train of manageable bunches, using current-technology rf cavities and power sources.[3,4] The same system can be used for both neutrino factory and collider scenarios.

The rf bunching naturally forms the beam for an initial cooling section. For the IDS neutrino factory design study, this uses a simple solenoidal focusing system with LiH absorbers that provides only transverse cooling (4-D phase-space cooling). More recently a 6-D initial cooling system (the "HFOFO Snake") using tilted solenoids and LiH wedge absorbers was developed,[5] and is currently considered somewhat superior, particularly for a collider scenario. The initial cooling concepts are discussed and compared.

## FRONT END OVERVIEW

The Front End concept presented here was generated for the Neutrino Factory design studies,[6] and subsequently extended and reoptimized for the Muon Accelerator Program (MAP) muon collider design studies. The Front End system takes the π's produced at the target, and captures and initiates cooling of the resulting decay μ's, preparing them for the μ accelerators. Fig. 2 shows an overview of the system, as recently developed for the MAP studies. In this figure, the transport past the target is separated into drift, buncher, rotator and cooling regions.

### Drift

The multi-GeV proton source produces short pulses of protons that are focused onto a target immersed in a high-field solenoid with an internal beam pipe radius $r_{sol}$. The proton bunch length is 1 to 3 ns rms (~5 to 15 ns full-width), $B_{sol}$ =20 T, and $r_{sol}$ = 0.075 m, at initial baseline parameters. Secondary particles are radially captured if they have a transverse momentum $p_T$ less than $\sim ecB_{sol}r_{sol}/2$ = 0.225 GeV/c. Downstream of the target solenoid the magnetic field is adiabatically reduced from 20T to 2T over ~14.75 m, while the beam pipe radius increases to ~0.25 m. This arrangement captures a secondary pion beam with a broad energy spread (~50 MeV to 400 MeV kinetic energy).

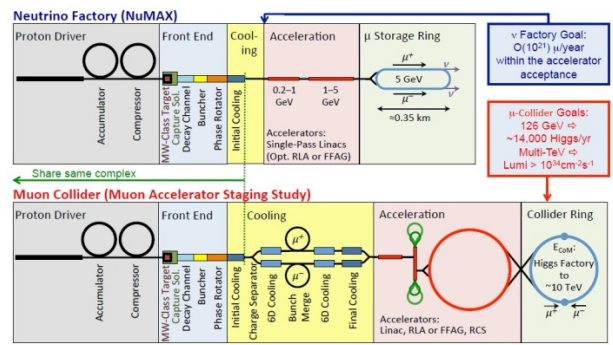

Figure 1: Block diagrams of neutrino factory and muon collider scenarios. The present paper discusses the "Front End" section from the Target to the Cooling/Acceleration, including "Initial Cooling". The same Front End can be used in both scenarios.

The initial proton bunch is relatively short, and as the secondary pions drift from the target they spread apart longitudinally, following: $c\tau(s) = s/\beta_z + c\tau_0$, where s is distance along the transport and $\beta_z = v_z/c$. Hence, downstream of the target, the pions and their daughter muons develop a position-energy correlation in the RF-free drift. In the MAP baseline, The total drift length is $L_D = 64.6$ m, and at the end of the decay channel there are about 0.2 muons of each sign per incident 8 GeV proton.

*Work supported by Contract No. De-AC02-07CH11359 with the U. S. Department of Energy.
#neuffer@fnal.gov

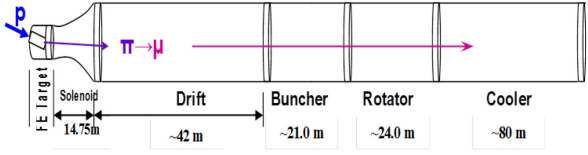

Figure 2A: Overview of the MAP front end, consisting of a target solenoid (20 T), a capture solenoid (20 T to 2.0T, 14.75 m), Drift section (42 m), rf Buncher (21 m), an energy-phase Rotator (24 m), with a Cooler (80 m).

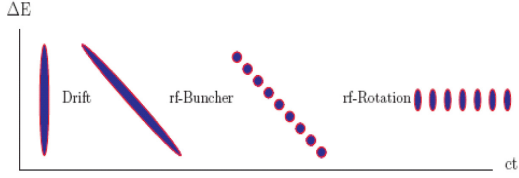

Figure 2B: Overview of longitudinal motion in the Front End. An intial muon distribution with large energy spread and small bunch length stretches to a distribution with an energy-position correlation in the Drift. The rf-Buncher forms the beam into a string of different energy bunches and the rf-Rotator moves the bunches to equal energies, forming a string of bunches for the downstream Cooler.

## RF Buncher

The drift channel is followed by a buncher section that uses rf cavities to form the muon beam into a train of bunches, and a phase-energy rotating section that decelerates the leading high-energy bunches and accelerates the later low-energy bunches to the same mean energy.[3] To determine the buncher parameters, we consider reference particles (0, N) at $P_0 = 250$ MeV/c and $P_N = 154$ MeV/c, with the intent of capturing muons from a large initial energy range (~50 to ~400 MeV). The rf frequency $f_{rf}$ and phase are set to place these particles at the center of bunches while the rf voltage increases along the transport. This requires that the rf wavelength $\lambda_{rf}$ increases, following:

$$N_B \lambda_{rf}(s) = N_B \frac{c}{f_{rf}(s)} = s\left(\frac{1}{\beta_N} - \frac{1}{\beta_0}\right)$$

where $s$ is the total distance from the production target, $\beta_1$ and $\beta_2$ are the velocities of the reference particles, and $N$ is an integer. For the baseline, $N$ is chosen to be 12, and the buncher length is 21m. Therefore, the rf cavities decrease in frequency from ~490 MHz ($\lambda_{rf} = 0.61$ m) to ~365 MHz ($\lambda_{rf} = 0.82$m) over the buncher length.

The initial geometry for rf cavity placement uses 2 0.25 m long cavities placed within 0.75 m long cells. The 2T solenoid field focusing of the decay region is continued through the Buncher and the Rotator. The rf gradient is increased along the Buncher, and the beam is captured into a string of bunches, each of them centered about a test particle position with energies determined by the $\delta(1/\beta)$ spacing from the initial test particle:

$$1/\beta_n = 1/\beta_0 + n\ \delta(1/\beta),$$

where $\delta(1/\beta) = (1/\beta_N - 1/\beta_0)/N$. In the initial design, the cavity gradients follow a linear increase along the buncher: $V'_{rf}(z) \approx 15\left(\dfrac{z}{L_{Bf}}\right)$ MV/m.

where $z$ is distance along the buncher and $L_{Bf}$ is the bucher section length. The gradual increase in voltage gradient enables a somewhat adiabatic capture of muons into separated bunches.

In practical implementation this linear ramp of varying-frequency cavities is approximated by a sequence of rf cavities that decrease in frequency along the 21 m length of the buncher. A total of 54 rf cavities are specified, with frequencies varying from 490 to 366 MHz, and rf gradients from 0 to 15 MV/m The number of different rf frequencies is limited to a more manageable 14 (~4 rf cavities per frequency). Table 1 lists the rf cavity requirements. At the end of the buncher, the beam is formed into a train of positive and negative bunches of different energies.

## Phase-Energy Rotator

In the rotator section, the rf bunch spacing between the reference particles is shifted away from the integer $N_B$ by an increment $\delta N_B$, and phased so that the high-energy reference particle is stationary and the low-energy one is uniformly accelerated to arrive at the high-energy at the end of the Rotator. For the MAP example, $\delta N_B = 0.05$ and the bunch spacing between the reference particles is $N_B + \delta N_B = 12.05$. The Rotator consists of 0.75 m long cells with 2 0.25 m rf cavities at 20 MV/m. The rf frequency of cavities decreases from 365 MHz to 326 MHz down the length of the 42 m long rotator region. The rotator uses 64 rf cavites of 16 different frequencies. (see Table 1) At the end of the rotator the rf frequency matches into the rf of the ionization cooling channel (325 MHz).

The constant solenoidal focusing field of 2 T is maintained throughout the rotator and is matched into an alternating solenoidal field for the cooler. The match is obtained by perturbations on the focusing coil currents of the first four cells of the cooler lattice.

## Cooler

The IDS baseline cooling channel design consists of a sequence of identical 1.5 m long cells (Fig. 2). Each cell contains two sets of two 0.25 m-long rf cavities, with 1.5cm thick LiH absorbers blocks at the ends of each cavity set (4 per cell) and a 0.25 m spacing between cavities, and alternating solenoidal focusing coils. The LiH provides the energy loss material for ionization cooling. The total length of the cooling section is ~75m (50 cells). Based on simulations, the cooling channel reduces the transverse emittances by a factor of 2.5.

The cells contain two solenoidal coils with the coils containing opposite sign currents. The coils produce an approximately sinusoidal variation of the magnetic field on axis in the channel with a peak value on-axis of ~2.8T, providing transverse focusing with $\beta_\perp \cong 0.8$m.

At the end of the cooler, simulations indicate that there are ~0.1 $\mu^+$ and $\mu^-$ per initial ~8 GeV proton within the projected acceptance of the downstream muon accelerator. (Particles with amplitudes $\varepsilon_{x,y} < $ ~0.03 m and londitudinal amplitudes $A_L <0.2$ m are considered to be within that acceptance.)

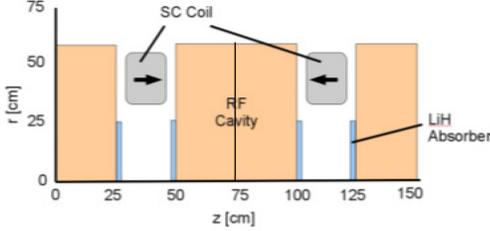

Figure 3. Layout of one period of the Alternating Solenoid cooling lattice, showing the alternating solenoid coils, the rf cavities and LiH absorbers.

Table 1. Summary of Front End rf requirements.

| Region | Number of Cavities | Number of frequencies | Frequencies [MHz] | Peak Gradient {MV/m] |
|---|---|---|---|---|
| Buncher | 54 | 14 | 490 to 366 | 0 to 15 |
| Rotator | 64 | 16 | 366 to 326 | 20 |
| Cooler | 200 | 1 | 325 | 25 |
| Total | 318 | 31 | 490 to 325 | 1700 MV |

*Rf breakdown, energy loss*

The baseline design requires the use of high-gradient rf (~325MHz, 15MV/m) in moderately high magnetic fields (~2T). Initial experiments and analysis showed that rf gradients could be limited in high magnetic fields, and it was uncertain that the baseline parameters could be achieved.[7, 8] More recent results using carefully prepared rf cavity surfaces have shown greater tolerance for magnetic fields. 800 MHz rf cavites operated at 20 MV/m within 5T magnetic fields in recent experiments.[9]

Experiments have also shown that gas-filled rf cavities suppress breakdown, even within high magnetic fields and the presence of rf accelerating beams. Variations of the Front End using gas-filled rf cavities were developed and found to obtain capture and acceptance approximately equal to vacuum cavity examples, provided the rf gradients are increased to compensate energy loss in the gas.[10] Operation of gas-filled rf has a complication in that beam loading by electrons from beam ionization can drain rf power; this can be moderated by adding a small fraction of electronegative dopant ($O_2$) to the $H_2$, to facilitate electron recombination.[11, 12]

The transport line from the target is designed to accept a maximal number of secondaries and therefore includes a large flux of particles that are not accepted into the final muon beams. These particles are mostly lost in the walls of the vacuum chamber (at up to 1 kW/m), potentially causing large activation. This includes a large flux of protons of all energies (up to the primary proton beam energy), as well as many pions and electrons. A chicane and absorber system following the target region was designed to localize losses to the target/chicane/absorber system, without greatly reducing downstream $\mu^+$ and $\mu^-$ acceptance.[13] Fig. 4 shows the front end modified with a chicane absorber; an additional length of ~30m is required.

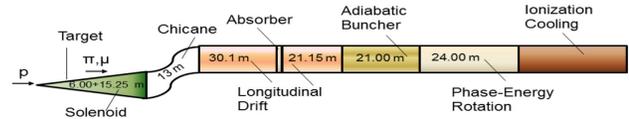

Figure 4. Layout of the Front End with a chicane and absorber.

## IMPROVED COOLING: THE "HFOFO SNAKE"

The "HFOFO Snake" was designed to obtain simultaneous transverse and longitudinal cooling, for both $\mu^+$ and $\mu^-$. The HFOFO snake is based on two principles: alternating solenoid focusing and resonant dispersion generation by a helical perturbation[5]. In a homogeneous longitudinal magnetic field the two transverse modes are the cyclotron and drift modes.[14] Ionization losses cool only the cyclotron mode. Changing the solenoid polarity exchanges the identity of the cyclotron and drift modes, so that in an alternating solenoid lattice both modes are damped. (This principle is also used in the initial baseline 2-D cooling channel.) In the HFOFO channel the solenoids are tilted in a periodic helical pattern, generating a helical closed orbit perturbation with dispersion. This perturbation introduces a path length increase with momentum (positive momentum compaction) through the absorbers which yields a longitudinal cooling effect. Wedges matched to the dispersion can give an additional cooling effect.

After some parameter variation and optimization, an HFOFO solution for the Front End was generated.[14] One period of the channel is shown in Fig. 5. It consists of:

• 6 Alternating solenoids (coil parameters: $L = 30$ cm, $R_{in} = 42$ cm, $R_{out} = 60$ cm) placed with a period of 70 cm along the axis. With current density 94.6 A/mm$^2$ the solenoids provide focusing with betatron phase advance ≈74°/step at a muon momentum of 230 MeV/c. To create a transverse magnetic field component, the solenoids are periodically inclined in rotating planes at $x \cdot \cos(\phi_k)+y \cdot \sin(\phi_k) = 0$, $\phi_k = \pi(1-2/N_s)(k+1)$, $k=1,2,…,N_s$. $N_s$ is the number of solenoids/period, $N_s = 6$ in the present case. The rotation angles $\phi_k$ are: $4\pi/3, 0, 2\pi/3, 4\pi/3, 0, 2\pi/3$; $\phi = 0$ corresponds to a tilt in the vertical plane. The chosen pitch angle of 2.5 mrad is too small to be visible in the figure.

- 6 Paired RF cavities ($f_{RF}$ = 325 MHz, $L$ = 2×25 cm, $E_{max}$ = 25 MV/m) filled with $H_2$ gas at a density that is 20% of liquid hydrogen, with Be windows. The radius and thickness of the Be windows are reduced in 3 steps along the channel: $R_w$ = 30 cm, $w$ = 0.12 mm (first 10 periods), $R_w$ = 25 cm, $w$ = 0.10 mm (next 10 periods) and $R_w$ = 20 cm, $w$ = 0.07 mm (last 10 periods). (Slab LiH absorbers could be used to replace the $H_2$, if vacuum-filled rf is preferred.)
- LiH wedge absorbers, providing additional longitudinal cooling. Although the momentum compaction factor of the lattice is positive, it is not sufficient for longitudinal cooling and wedges must be added. The wedge angle smoothly varies along the channel from 0.17 rad to 0.20 rad. The tip of the wedge just intercepts the channel axis so the muons on the equilibrium orbit (see Fig. 5) traverse no more than 0.3 mm of LiH.

An important feature of the design is that $\mu^-$ in solenoids 4, 5, 6 see exactly the same forces as $\mu^+$ in solenoids 1, 2, 3 and vice versa, so that $\mu^-$ and $\mu^+$ orbits have exactly the same form, with a longitudinal shift by a half period (three solenoids), and are not mirror-symmetric as one might expect. This allows us to find an orientation of the wedge absorbers (with periodicity = 2) such that they provide longitudinal cooling for both $\mu^-$ and $\mu^+$. The complete cooling channel contains 30 periods for a total length of 126 m.

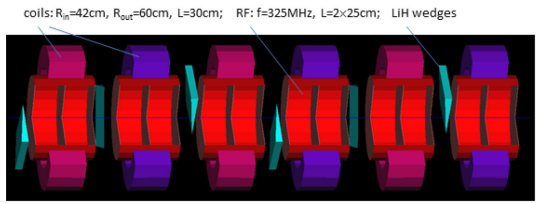

Figure 5. Layout of one period of the HFOFO lattice, showing the alternating solenoid coils (violet and blue), the rf cavities (red) and wedge LiH absorbers (cyan).

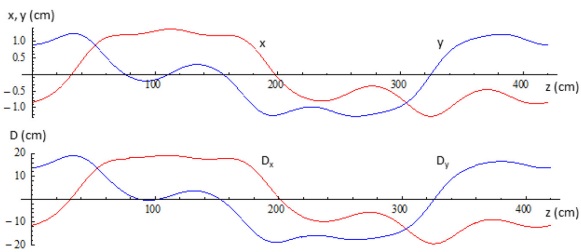

Figure 6. Closed orbit(above) and dispersion (below) in one period of the HFOFO lattice.

The closed orbit through the period plus the dispersion (for $\mu^+$) are shown in figure 6. The closed orbit approximately follows a circle of radius ~1.2cm, and the dispersion follows a ~20cm amplitude circle through the helical orbit. The normal mode tunes of the lattice are ($\nu_1$, $\nu_2$, $\nu_3$) = (1.23+0.01i, 1.238+0.0036i, 0.189+0.0049i), where the first two modes are transverse and the last is predominantly longitudinal; the imaginary part indicates the damping from ionization cooling.

The HFOFO channel was matched in closed orbit and betatron functions to the output of the Front End Rotator, using perturbations of the first 9 HFOFO solenoid strengths and tilts. In the HFOFO, the central momentum is gradually reduced from the ~250 MeV/c of the Rotator to ~210 MeV/c at the end of the HFOFO cooler.

Beam from the Front End Rotator was matched into the HFOFO cooler and tracked using the G4BeamLine simulation code. Results of the simulation are presented in Figure [7]. Both $\mu^+$ and $\mu^-$ beams are cooled. The rms transverse emittance $\varepsilon_t$ =($\varepsilon_1 \varepsilon_2$)$^{1/2}$ is reduced from ~16 mm to ~2.6mm, which is a factor of ~6. Longitudinal emittance is reduced from ~24 mm to ~7.4 mm, about a factor of 3. Beam survival of the core muon beams (from decay and aperture losses) was ~70%.

The transverse cooling is greater than that of the previous 2-D cooling system. This is in part due to the stronger focussing and longer channel length. Initial losses are ~10% greater, partly from the more complicated 6-D phase space match, and partly from aperture losses in adding the helical orbit to the transverse motion. However the additional longitudinal cooling enables the use of the longer channel and provides a better match into downstream systems, which are likely to have more limited longitudinal acceptances.

Because of the superior cooling and the better match into downstream systems, The HFOFO snake is presently the preferred initial cooling system.

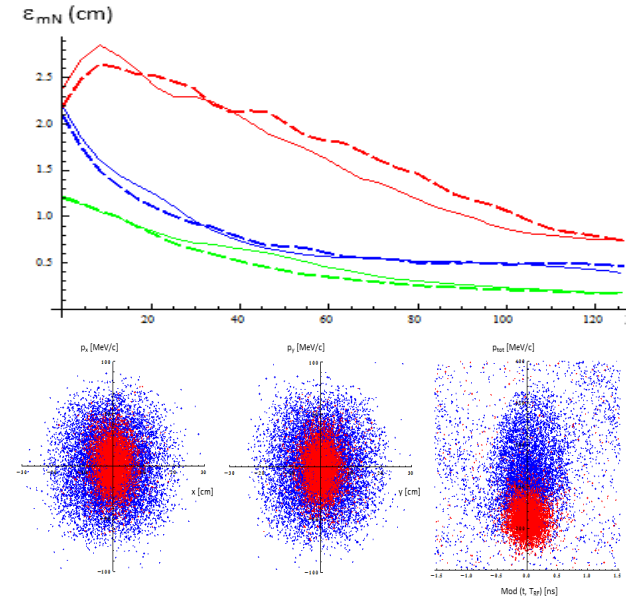

Figure 7. Results of beam simulations of cooling. (Above) Evolution of the eigenemittances ($\varepsilon_1$, $\varepsilon_2$, $\varepsilon_3$) through the 124m channel. (solid lines are $\mu^+$, dashed are $\mu^-$) (Below) Phase space distributions of the initial $\mu$+ beam (blue) and of the cooled beam in the exit solenoid (red). (x-$p_x$, y-$p_y$, p-ct) All bunches were projected onto the same RF bucket in the last plot. No cuts applied.

## CONCLUSION

The results here present a baseline design for the muon Front End as developed by the MAP collaboration. It would provide a large-acceptance and capture system that could provide an intense source of bunched and cooled μ+ and μ-. The example presented is certainly not a final, optimized design and the research has suggested many possibilities for future improvement and modification.
For example, the 2T baseline focussing is not optimum; stronger focusing of ~3T should provide more capture.

The present channel was developed for muon collider/neutrino factory applications. The same methods could also be adapted to obtain intense low-energy muon sources.

## ACKNOWLEDGMENT

We thank M. Palmer, D. Stratakis, C. Rogers, P. Snopok, V. Shiltsev, and the MAP collaboration for helpful contributions and support.

## REFERENCES

[1] M. Appollonio et al., "Accelerator Concept for Future Neutrino Facilities", RAL-TR-2007-23, *JINST* **4** P07001 (2009).

[2] J. P. Delahaye et al., *Enabling Intensity and Energy Frontier Science with a Muon Accelerator Facility in the U. S.*, FERMILAB-CONF-13-307-APC, (2013).

[3] D. Neuffer and A. Van Ginneken, Proc. PAC 2001, Chicago IL p.2029 (2001).

[4] J. S. Berg *et al.*, "Cost-effective Design for a Neutrino Factory", **Phys. Rev. STAB 9**, 011001(2006).

[5] Y. Alexahin, "Helical FOFO snake for 6D ionization cooling of muons", AIP Conf. Proc. 1222 (2010), pp. 318-323. see also Y. Alexahin, "Helical FOFO snake for initial 6D cooling of muons", **ICFA Beam Dyn. Newsletter 65** (2014) pp. 49-54.

[6] C. T. Rogers et al., "Muon Front End for a Neutrino Factory", **Phys. Rev. STAB 16**, 040104(2013).

[7] A. Moretti et al., "Effects of high solenoidal magnetic fields on rf accelerating cavities", **Phys. Rev. STAB 8**, 072001 (2005).

[8] R. B. Palmer et al., "rf breakdown with external magnetic fields in 201 and 805 MHz cavities", **Phys. Rev. STAB 12**, 031002 (2009).

[9] D. Bowring et al. "rf Breakdown of 805 MHz Cavities in Strong Magnetic Fields", Proc. IPAC2015, Richmond VA, p. 53 (2015).

[10] D Stratakis and D. Neuffer, "Utilizing gas filled cavities for the generation of an intense muon source", Proc. IPAC2015, Richmond VA USA, p. 2830 (2015).

[11] B. Freemire et al., "High-Pressure rf Cavities for Use in a Muon Cooling Channel,", Proc. PAC 2013, Pasadena, CA USA, p. 419 (2013).

[12] J. Ellison et al., "Beam-plasma Effects in Muon Ionization Cooling Lattices", Proc. of IPAC2015, Richmond, VA USA, p. 2649 (2015)

[13] C. Rogers, D. Neuffer, and P. Snopok, "Control of beam losses in the Front End for the Neutrino Factory", Proc. IPAC2012, New Orleans LA, p, 223 (2012).

[14] A. Burov, Y. Derbenev, S. Nagaitsev, **Phys. Rev. E 66**, 016503 (2002).

[15] Y. Alexahin, "H2 Gas-Filled Helical FOFO Snake for Initial 6D Ionization Cooling of Muons", MAP-doc-4377, http://map-docdb.fnal.gov/cgi-bin/ShowDocument?docid=4377, May 2014.